\documentclass[twocolumn,prl]{revtex4}

\usepackage{amsmath,amssymb,bm}
\usepackage{graphicx}
\usepackage{hyperref}

\newcommand{\fig}[1]{Fig.~\ref{#1}}

\newcommand{\rim}     {\rho_\textrm{imag}}
\newcommand{\Vim}    {V_\textrm{imag}}
\newcommand{\Cs}     {${}^{133}\textrm{Cs}$}
\newcommand{\Li}[1]  {${}^{#1}\textrm{Li}$}

\newcommand{\rec}    {\alpha_\textrm{rec}}

\begin{document}

\title{Three-body recombination at finite energy within an optical model}

\author{P. K. S{\o}rensen, D. V. Fedorov, A. S. Jensen and N. T. Zinner}
\affiliation{Department of Physics and Astronomy - Aarhus University, Ny
Munkegade, bygn. 1520, DK-8000 {\AA}rhus C, Denmark}

\begin{abstract}
We investigate three-boson recombination of equal mass systems as function
of (negative) scattering length, mass, finite energy, and finite
temperature. 
An optical model with an imaginary potential at short distance
reproduces experimental recombination data and allows 
us to provide a simple parametrization
of the recombination rate as function of scattering length and energy.
Using the
two-body van der Waals length as unit we find that the imaginary potential 
range and also the potential depth agree to within thirty percent for
Lithium and Cesium atoms. As opposed to recent studies suggesting 
universality of the threshold for bound state formation, our results
suggest that the recombination process itself could have universal 
features.
\end{abstract}
\maketitle

The quantum mechanical three-body problem has a
long and rich history. A milestone that has spurred great interest came
in the early 1970ties when Efimov \cite{efi1970} found that three bosonic
particles with short-range interactions can have an infinite number of
three-body bound states when each two-body subsystem has a bound state
with exactly zero binding energy. This intriguing possibility was pursued
for many years in the context of nuclear physics \cite{nie01,jen2004,bra2006},
but its first experimental signature came using an ultra cold atomic gas
of Cesium atoms in 2006 \cite{kraemer2006}. The use of tunable atomic
interactions through the principle of Feshbach resonance \cite{chin2010}
has enabled the study of these so-called Efimov three-body states in a
number of different atomic alkali systems at low temperature 
\cite{ottenstein2008, pollack2009, zaccanti2009, gross2009, huckans2009,
williams2009, lompe2010a, gross2010, lompe2010b, nakajima2010,
nakajima2011, wild2012, machtey2012a, machtey2012b, knoop2012, hulet2013, roy2013}.
While Efimov states are intrinsically low-energy spatially extended
so-called universal states, the overall energy scale was previously
believed to be given by short-range physics through a so-called three-body
parameter, $\Lambda$ \cite{bra2006}. A highly suggestive relation between
$\Lambda$ and the inter-atomic two-body van der Waals interactions was
nevertheless recently suggested \cite{berninger2011,naidon2011}. A number
of theoretical papers linking the short- and long-range energy scales of
universal three-body physics soon followed
\cite{chin2011,wang2012,sorensen2012,schmidt2012,naidon2012} that hit at a
universal short-range barrier in the effective three-body potential
determined solely by parameters of the two-body inter-atomic potential.
This has been interpreted as universality of the three-body parameter
itself.

\begin{figure}[ht!]
  \centering
\begingroup
  \makeatletter
  \providecommand\color[2][]{%
    \GenericError{(gnuplot) \space\space\space\@spaces}{%
      Package color not loaded in conjunction with
      terminal option `colourtext'%
    }{See the gnuplot documentation for explanation.%
    }{Either use 'blacktext' in gnuplot or load the package
      color.sty in LaTeX.}%
    \renewcommand\color[2][]{}%
  }%
  \providecommand\includegraphics[2][]{%
    \GenericError{(gnuplot) \space\space\space\@spaces}{%
      Package graphicx or graphics not loaded%
    }{See the gnuplot documentation for explanation.%
    }{The gnuplot epslatex terminal needs graphicx.sty or graphics.sty.}%
    \renewcommand\includegraphics[2][]{}%
  }%
  \providecommand\rotatebox[2]{#2}%
  \@ifundefined{ifGPcolor}{%
    \newif\ifGPcolor
    \GPcolorfalse
  }{}%
  \@ifundefined{ifGPblacktext}{%
    \newif\ifGPblacktext
    \GPblacktexttrue
  }{}%
  \let\gplgaddtomacro\g@addto@macro
  \gdef\gplbacktext{}%
  \gdef\gplfronttext{}%
  \makeatother
  \ifGPblacktext
    \def\colorrgb#1{}%
    \def\colorgray#1{}%
  \else
    \ifGPcolor
      \def\colorrgb#1{\color[rgb]{#1}}%
      \def\colorgray#1{\color[gray]{#1}}%
      \expandafter\def\csname LTw\endcsname{\color{white}}%
      \expandafter\def\csname LTb\endcsname{\color{black}}%
      \expandafter\def\csname LTa\endcsname{\color{black}}%
      \expandafter\def\csname LT0\endcsname{\color[rgb]{1,0,0}}%
      \expandafter\def\csname LT1\endcsname{\color[rgb]{0,1,0}}%
      \expandafter\def\csname LT2\endcsname{\color[rgb]{0,0,1}}%
      \expandafter\def\csname LT3\endcsname{\color[rgb]{1,0,1}}%
      \expandafter\def\csname LT4\endcsname{\color[rgb]{0,1,1}}%
      \expandafter\def\csname LT5\endcsname{\color[rgb]{1,1,0}}%
      \expandafter\def\csname LT6\endcsname{\color[rgb]{0,0,0}}%
      \expandafter\def\csname LT7\endcsname{\color[rgb]{1,0.3,0}}%
      \expandafter\def\csname LT8\endcsname{\color[rgb]{0.5,0.5,0.5}}%
    \else
      \def\colorrgb#1{\color{black}}%
      \def\colorgray#1{\color[gray]{#1}}%
      \expandafter\def\csname LTw\endcsname{\color{white}}%
      \expandafter\def\csname LTb\endcsname{\color{black}}%
      \expandafter\def\csname LTa\endcsname{\color{black}}%
      \expandafter\def\csname LT0\endcsname{\color{black}}%
      \expandafter\def\csname LT1\endcsname{\color{black}}%
      \expandafter\def\csname LT2\endcsname{\color{black}}%
      \expandafter\def\csname LT3\endcsname{\color{black}}%
      \expandafter\def\csname LT4\endcsname{\color{black}}%
      \expandafter\def\csname LT5\endcsname{\color{black}}%
      \expandafter\def\csname LT6\endcsname{\color{black}}%
      \expandafter\def\csname LT7\endcsname{\color{black}}%
      \expandafter\def\csname LT8\endcsname{\color{black}}%
    \fi
  \fi
  \setlength{\unitlength}{0.0500bp}%
  \begin{picture}(5040.00,3528.00)%
    \gplgaddtomacro\gplbacktext{%
      \csname LTb\endcsname%
      \put(1078,704){\makebox(0,0)[r]{\strut{}-0.2}}%
      \put(1078,1024){\makebox(0,0)[r]{\strut{}-0.15}}%
      \put(1078,1344){\makebox(0,0)[r]{\strut{}-0.1}}%
      \put(1078,1664){\makebox(0,0)[r]{\strut{}-0.05}}%
      \put(1078,1984){\makebox(0,0)[r]{\strut{} 0}}%
      \put(1078,2303){\makebox(0,0)[r]{\strut{} 0.05}}%
      \put(1078,2623){\makebox(0,0)[r]{\strut{} 0.1}}%
      \put(1078,2943){\makebox(0,0)[r]{\strut{} 0.15}}%
      \put(1078,3263){\makebox(0,0)[r]{\strut{} 0.2}}%
      \put(1691,484){\makebox(0,0){\strut{}$\rim$}}%
      \put(1210,484){\makebox(0,0){\strut{} 0}}%
      \put(1897,484){\makebox(0,0){\strut{} 1}}%
      \put(2583,484){\makebox(0,0){\strut{} 2}}%
      \put(3270,484){\makebox(0,0){\strut{} 3}}%
      \put(3956,484){\makebox(0,0){\strut{} 4}}%
      \put(4643,484){\makebox(0,0){\strut{} 5}}%
      \put(176,1983){\rotatebox{-270}{\makebox(0,0){\strut{}$V(\rho)\dfrac{ma^2}{\hbar^2}$}}}%
      \put(2926,154){\makebox(0,0){\strut{}$\rho/|a|$}}%
      \put(2755,2828){\makebox(0,0)[l]{\strut{}$f(\rho) = He^{-ik\rho}+Ge^{ik\rho}$}}%
    }%
    \gplgaddtomacro\gplfronttext{%
      \csname LTb\endcsname%
      \put(2732,1537){\makebox(0,0)[l]{\strut{}$V(\rho)$}}%
      \csname LTb\endcsname%
      \put(2732,1317){\makebox(0,0)[l]{\strut{}Bound state}}%
      \csname LTb\endcsname%
      \put(2732,1097){\makebox(0,0)[l]{\strut{}Resonance}}%
      \csname LTb\endcsname%
      \put(2732,877){\makebox(0,0)[l]{\strut{}$V_\textrm{imag}$}}%
    }%
    \gplbacktext
    \put(0,0){\includegraphics{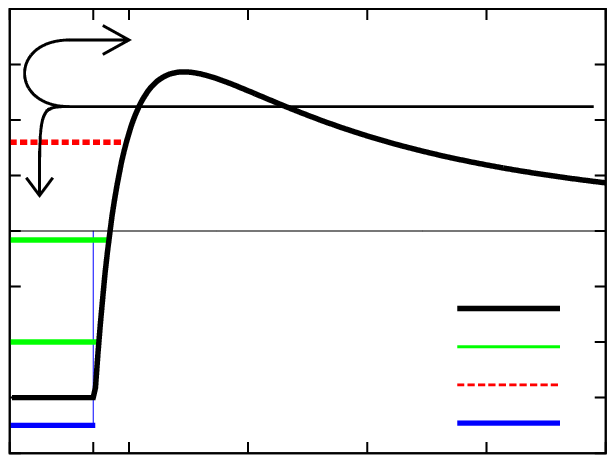}}%
    \gplfronttext
  \end{picture}%
\endgroup
  \caption{Three-body model potential for negative scattering length $a$ 
    as function of hyperradius $\rho$.
    The actual $\rim$ used in calculations is much smaller than
    illustrated. The potential drops as $1/\rho^2$ at large distances and
    has a constant (complex) value for $\rho<\rim$. The split arrow
    indicates that amplitiude is both reflected and absorbed via the    
    complex potential. Green lines illustrate bound
    states in the potential while the green dashed line indicates a
    resonance.}
  \label{figure:1}
\end{figure}

Here we present an entirely different approach to the problem of
universality of three-body physics. The universality suggested in
Ref.~\cite{berninger2011} was found in a region of parameter space where
there are no weakly-bound two-body subsystems. There the observed recombination
loss which is the most common signature of Efimov states takes place at
short distance through decay into deeply bound two-body molecular states
\cite{bra2006}. The problem then becomes how to match the long-distance
universal three-body behaviour to a short-distance region capable of
describing absorption due to three-body loss. In the present paper
we construct a physically transparent model that uses a minimal number of
parameters and does {\it not} require any short-range barrier. Instead we
employ an imaginary potential inspired by optical models used in other
areas of physics.

In Fig.~\ref{figure:1} we show the effective potential for the
three-body system. At long-distance there is a repulsive inverse-square which
is separated by a barrier from an attractive inverse-square at intermediate 
distance, and at short
distance an imaginary potential. The short-distance part may then be
related to the two-body scale at which the decay to deeply bound states
occurs. Below we show that this type of model describes experimental data
for loss rates at different collision energies and/or finite temperatures
extremely well. In addition, we demonstrate that when the imaginary part
of the potential causing the loss is implemented in a (small) region of
space the decay parameter, $\gamma$, depends on
both scattering length, $a$, and on the collision energy. It also shows
resonant features that obey appropriate Efimov scaling relations. This is
in contrast to many previous models
\cite{nielsen1999,esry1999,bed2000,braaten2001} where an analogous parameter 
(typically denoted $\eta$) is either found to be or assumed to be 
independent of energy and $a$.

A system of three bosonic particles of equal mass $m$ at position
$\bm r_i$, $\bm r_j$ and $\bm r_k$ is treated using hyperspherical
coordinates defined by $\bm x_i=\left(\bm r_j-\bm r_k\right)/\sqrt{2}$ and
$\bm y_i=\sqrt{\tfrac{2}{3}}\left(\bm r_i-\left(\bm r_j+\bm r_k\right)/2\right)$,
where the Jacobi indices $\{i,j,k\}$ are cyclic permutations of
$\{1,2,3\}$. The hyperradius $\rho^2=|\bm x_i|^2+|\bm y_i|^2$ is
independent of the choice of index. One hyperangle is defined as
$\alpha_i=\tan^{-1} \frac{|\bm x_i|}{|\bm y_i|}$ and the remaining four by
the directions of $\bm x_i$ and $\bm y_i$. 
We use the hyperspherical adiabatic approximation, the Faddeev
decomposition and zero-range potentials to describe the interactions
between the atoms. All methods are documented in detail in Ref.~\cite{twochannelFeshbach}.
This yields a hyperradial differential equation
\begin{equation}
  \left(-\frac{d^2}{d\rho^2}+\frac{\nu^2(\rho)-1/4}{\rho^2}-
  \frac{2mE}{\hbar^2}\right)f(\rho)=0\;,
  \label{eq:2}
\end{equation}
where $\nu(\rho)$ is determined implicitly by
\begin{equation}
  \frac{\nu\cos\left(\dfrac{\nu\pi}{2}\right)-
    \dfrac{8}{\sqrt{3}}\sin\left( \dfrac{\nu\pi}{6}\right)}
    {\sin\left(\dfrac{\nu\pi}{2}\right)} = \sqrt{2}\frac{\rho}{a}\;,
  \label{eq:3}
\end{equation}
where $a$ is the common scattering length between each pair of particles.
For negative scattering lengths the potential
$V(\rho)=\frac{\nu^2(\rho)-1/4}{\rho^2}$ has a barrier region which the
wave-function must penetrate, see \fig{figure:1}. The limits of $\nu^2$ is
$4$ and $-1.012$ for large and small $\rho$, respectively. The potential
is zero when $\nu=1/2$ corresponding to $\rho=0.84a$, and the barrier
maximum is at $\rho=1.46a$ with the peak height $E_B(a)=0.143 \hbar^2/(ma^2)$.
The divergence due to $1/\rho^2$ as $\rho$ vanishes provides the Efimov
scaling of $22.7$ for bound or resonance states at zero energy. We select
the range of these three-body states by continuing the real potential as a
constant with the same value between $0$ and the cut-off $\rho_\textrm{imag}$.
Furthermore, we add a constant short-range imaginary potential,
$V(\rho<\rim)=\Vim$, acting as a probability sink which
models recombination to deep dimers, see \fig{figure:1}. This is in stark
contrast to the usual regularization cut-off method where the potential is
set to infinity for $\rho$-values below some small cut-off value. This 
is important as it means that we do not use repulsive core at short
distance as in many previous studies. 
In this paper we use a completely different strategy where we avoid
a three-body cut-off but obtain a model for the loss that matches 
experimental data as we show below. The imaginary potential reflects
short-distance two-body effects coming from 
the neglected Hilbert space with deeply bound dimers. This can
formally be done through Feshbach reaction theory that facilitates 
reduction to a smaller active Hilbert space.

We calculate the three-body recombination rate $\rec$ for negative
scattering length using the radial equation Eq.\eqref{eq:2} directly. The
rate is defined as $\dot n=-\rec n^3$ where $n$ is the particle density.
At large hyperradii the wave-function is decomposed into incoming and
outgoing plane waves $f(\rho)=He^{-ik\rho}+Ge^{ik\rho}$ with
$k^2=2mE/\hbar^2$. The probability of recombination is simply
$R=1-|G/H|^2$. The loss of probability is quantified using a complex
phase-shift between incoming and outgoing waves
\begin{equation}
  G=e^{2i(\theta+i\gamma)}H \; .
  \label{eq:4}
\end{equation}
The recombination probability, $R$, is proportional to the absorption
cross section and we write $R =1-e^{-4\gamma}$ where $\gamma$
is the decay parameter. The recombination rate is obtained
using the method of Ref.~\cite{twochannelFeshbach}, and given by
\begin{equation}
  \rec=4(2\pi)^23\sqrt3\frac{\hbar^5}{m^3}
 \frac{R}{E^2}\;.
  \label{eq:5}
\end{equation}
This is obtained after dividing $R$ by $E^2$, that is the
initial three-body phase space, which also is responsible for the
well defined limit of zero energy where $\gamma\propto E^2$. The
usual recombination rate for zero energy is then found from
Eq.~\eqref{eq:5} by letting $E \rightarrow 0$.

For absorption to take place the particles must cross or tunnel through
the barrier and enter the region of finite imaginary potential. Barriers
like in \fig{figure:1} may exhibit resonant behaviour around specific
energies characterized by an abrupt change by $\pi$ of the real part of
the phase shift, $\theta$. A resonance, or bound state, may appear at zero
energy with the corresponding scattering length denoted as ${a^{(-)}}$. At
a resonance the probability for penetrating into the absorptive potential region
is substantially increased. Consequently, the recombination rate is also
substantially increased when three-body ($E$) and resonance ($E_0$)
energies coincide.

This discussion in terms of scattering resonances strongly
indicates that $\rec$ can be parametrized by a Breit-Wigner distribution,
that is
\begin{equation}
  \rec(a,E)
  =\frac{ 4(2\pi)^23\sqrt3 K}{\left[E-E_0(a)\right]^2
 +\frac{1}{4}\Gamma^2\left(a\right)} \frac{\hbar^5}{m^3}\;,
  \label{eq:6}
\end{equation}
where the numerical factor is chosen for easy comparison to
Eq.~\eqref{eq:5}. This
expression exhibits the physical interpretation of tunnelling through the
barrier and subsequently subject to absorption and reflection at short
distance. The dimensionless constant $K$ describes
absorption and depends only on the imaginary potential. 

We can give an alternative argument for the form of the recombination 
rate that is based on the WKB formulation which has been succesfully
used for few-body recombination in Refs.~\cite{twochannelFeshbach,peder2013}.
The WKB tunneling probability for given energy, $E$, through the
$\nu^2/\rho^2$ barrier in Eq.\eqref{eq:2} is easily calculated. We have
here added the Langer correction effectively removing the term $-1/4$ as
required for a semi-classical calculation. The classical turning points
are then $\rho =2/\sqrt{2 m E/\hbar^2}$ and $\rho \approx a$, where we use
the large-distance limit of $\nu=2$ and that the potential strongly
decreases inside the barrier. The corresponding
probability depends on $\exp(-2S) \propto E^2 a^4$ (with $S$ is the 
action integral \cite{peder2013}), and scales precisely
with $E^2$ where the power of 2 is directly related to the
large $\rho$ limit with $\nu =2$. This is dictated by the three-body phase
space dependence on energy. The second order WKB expression is 
$1/(1+\exp(2S))$ \cite{froman2002} 
which improves the first order WKB, $\exp(-2S)$, when
$S$ approaches zero for energies close to the barrier height. Thus, the
WKB tunnelling approximation also leads to the Breit-Wigner
parametrization in Eq.~\eqref{eq:6}. The resonance
position cannot, however, be determined from barrier properties.

Combining cross section, tunneling probability, and absorption potential
yields an interpretation in terms of a two-step process;
tunneling at large distance followed by absorption/reflection at short
distance. When the barrier is negligible compared to the energy, all
resonance structures are completely smeared out. Furthermore, as seen
from Eq.~\eqref{eq:6}, the recombination rate reaches an upper bound
inversely proportional to the energy. This is known as the unitarity
limit \cite{dincao2004}.

The objective is now to determine the parameters, $\rim$ and $\Vim$, of the
imaginary potential, preferentially to fit experimental data by full
numerical calculation. In general $\rim$ sets the location of
recombination peaks as function of scattering length while $\Vim$
determines the overall shape of these peaks. Both parameters are
short-range parameters reflecting that recombination requires all three
particles to be close to each other for recombination to occur. However,
the final states are bound dimers with one free particle. In fact,
it has recently been argued
that the short-range three-body cut-off is determined
by the short-range two-body repulsion. That argument involves the 
threshold for the appearance of the first Efimov three-body 
state out of the three-atom continuum \cite{berninger2011}. While this
point is typically determined by recombination loss peaks, previous
theoretical models have only discussed the spectrum and not the loss.

\begin{figure}[ht!]
  \centering
\begingroup
  \makeatletter
  \providecommand\color[2][]{%
    \GenericError{(gnuplot) \space\space\space\@spaces}{%
      Package color not loaded in conjunction with
      terminal option `colourtext'%
    }{See the gnuplot documentation for explanation.%
    }{Either use 'blacktext' in gnuplot or load the package
      color.sty in LaTeX.}%
    \renewcommand\color[2][]{}%
  }%
  \providecommand\includegraphics[2][]{%
    \GenericError{(gnuplot) \space\space\space\@spaces}{%
      Package graphicx or graphics not loaded%
    }{See the gnuplot documentation for explanation.%
    }{The gnuplot epslatex terminal needs graphicx.sty or graphics.sty.}%
    \renewcommand\includegraphics[2][]{}%
  }%
  \providecommand\rotatebox[2]{#2}%
  \@ifundefined{ifGPcolor}{%
    \newif\ifGPcolor
    \GPcolorfalse
  }{}%
  \@ifundefined{ifGPblacktext}{%
    \newif\ifGPblacktext
    \GPblacktexttrue
  }{}%
  \let\gplgaddtomacro\g@addto@macro
  \gdef\gplbacktext{}%
  \gdef\gplfronttext{}%
  \makeatother
  \ifGPblacktext
    \def\colorrgb#1{}%
    \def\colorgray#1{}%
  \else
    \ifGPcolor
      \def\colorrgb#1{\color[rgb]{#1}}%
      \def\colorgray#1{\color[gray]{#1}}%
      \expandafter\def\csname LTw\endcsname{\color{white}}%
      \expandafter\def\csname LTb\endcsname{\color{black}}%
      \expandafter\def\csname LTa\endcsname{\color{black}}%
      \expandafter\def\csname LT0\endcsname{\color[rgb]{1,0,0}}%
      \expandafter\def\csname LT1\endcsname{\color[rgb]{0,1,0}}%
      \expandafter\def\csname LT2\endcsname{\color[rgb]{0,0,1}}%
      \expandafter\def\csname LT3\endcsname{\color[rgb]{1,0,1}}%
      \expandafter\def\csname LT4\endcsname{\color[rgb]{0,1,1}}%
      \expandafter\def\csname LT5\endcsname{\color[rgb]{1,1,0}}%
      \expandafter\def\csname LT6\endcsname{\color[rgb]{0,0,0}}%
      \expandafter\def\csname LT7\endcsname{\color[rgb]{1,0.3,0}}%
      \expandafter\def\csname LT8\endcsname{\color[rgb]{0.5,0.5,0.5}}%
    \else
      \def\colorrgb#1{\color{black}}%
      \def\colorgray#1{\color[gray]{#1}}%
      \expandafter\def\csname LTw\endcsname{\color{white}}%
      \expandafter\def\csname LTb\endcsname{\color{black}}%
      \expandafter\def\csname LTa\endcsname{\color{black}}%
      \expandafter\def\csname LT0\endcsname{\color{black}}%
      \expandafter\def\csname LT1\endcsname{\color{black}}%
      \expandafter\def\csname LT2\endcsname{\color{black}}%
      \expandafter\def\csname LT3\endcsname{\color{black}}%
      \expandafter\def\csname LT4\endcsname{\color{black}}%
      \expandafter\def\csname LT5\endcsname{\color{black}}%
      \expandafter\def\csname LT6\endcsname{\color{black}}%
      \expandafter\def\csname LT7\endcsname{\color{black}}%
      \expandafter\def\csname LT8\endcsname{\color{black}}%
    \fi
  \fi
  \setlength{\unitlength}{0.0500bp}%
  \begin{picture}(5040.00,3528.00)%
    \gplgaddtomacro\gplbacktext{%
      \csname LTb\endcsname%
      \put(726,704){\makebox(0,0)[r]{\strut{}-0.1}}%
      \put(726,1024){\makebox(0,0)[r]{\strut{} 0}}%
      \put(726,1344){\makebox(0,0)[r]{\strut{} 0.1}}%
      \put(726,1664){\makebox(0,0)[r]{\strut{} 0.2}}%
      \put(726,1984){\makebox(0,0)[r]{\strut{} 0.3}}%
      \put(726,2303){\makebox(0,0)[r]{\strut{} 0.4}}%
      \put(726,2623){\makebox(0,0)[r]{\strut{} 0.5}}%
      \put(726,2943){\makebox(0,0)[r]{\strut{} 0.6}}%
      \put(726,3263){\makebox(0,0)[r]{\strut{} 0.7}}%
      \put(858,484){\makebox(0,0){\strut{} 0}}%
      \put(1489,484){\makebox(0,0){\strut{} 20}}%
      \put(2120,484){\makebox(0,0){\strut{} 40}}%
      \put(2751,484){\makebox(0,0){\strut{} 60}}%
      \put(3381,484){\makebox(0,0){\strut{} 80}}%
      \put(4012,484){\makebox(0,0){\strut{} 100}}%
      \put(4643,484){\makebox(0,0){\strut{} 120}}%
      \put(2750,154){\makebox(0,0){\strut{}$|\Vim|\dfrac{m\rim^2}{\hbar^2}$}}%
    }%
    \gplgaddtomacro\gplfronttext{%
      \csname LTb\endcsname%
      \put(3656,3090){\makebox(0,0)[r]{\strut{}K}}%
      \csname LTb\endcsname%
      \put(3656,2870){\makebox(0,0)[r]{\strut{}A}}%
      \csname LTb\endcsname%
      \put(3656,2650){\makebox(0,0)[r]{\strut{}$\delta$}}%
      \csname LTb\endcsname%
      \put(3656,2430){\makebox(0,0)[r]{\strut{}B}}%
      \csname LTb\endcsname%
      \put(3656,2210){\makebox(0,0)[r]{\strut{}$\beta$}}%
      \csname LTb\endcsname%
      \put(3656,1990){\makebox(0,0)[r]{\strut{}$1-e^{-4\gamma}$}}%
    }%
    \gplbacktext
    \put(0,0){\includegraphics{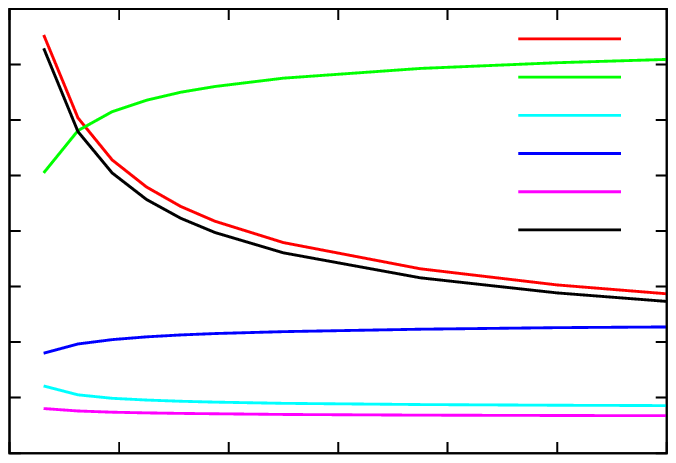}}%
    \gplfronttext
  \end{picture}%
\endgroup
  \caption{The parameters of Eqs.~\eqref{eq:7} and \eqref{eq:8} as
    functions of the strength of the imaginary square-well potential. All quantities are
    dimensionless.}
  \label{figure:3}
\end{figure}

The next step in the parametrization is to find $K$, $E_0$, and $\Gamma$.
The choice of factors in Eq.~\eqref{eq:6} immediately gives the high-energy
limit, $K\to 1-\exp(-4\gamma)$, where $E$ has to be large compared to other
terms in the denominator of Eq.~\eqref{eq:6}. The peaks appearing when
$a=a^{(-)}$ at intervals of $\exp(\pi/s_0) \approx 22.7$ ($s_0=1.00624$) require
corresponding periodicity. Both $\Gamma$ and $E_0$ can be parametrized
by
\begin{align}
  \Gamma^2(a)\frac{m^2a^4}{\hbar^4}&=A\sin^2\!
    \left[s_0\ln\left(\frac{a}{a^{(-)}}\right)\right]+\delta\;,
  \label{eq:7}\\
  E_0(a)\frac{ma^2}{\hbar^2}&=B\sin^{\phantom{2}}\!
    \left[s_0\ln\left(\frac{a}{a^{(-)}}\right)\right]+\beta\;,
  \label{eq:8}
\end{align}
where $A,B,\beta,\delta$ are constants that weakly depend on the
imaginary potential. This form ensures that both the $a^4$-rule and the
Efimov scaling are obeyed with periodic $22.7$ peak-recurrence for given
$a=a^{(-)}$.
In~\fig{figure:3} we plot the parameters for each
imaginary strength, $|\Vim|m\rim^2/\hbar^2$, obtained by fitting Eqs.~\eqref{eq:6},
\eqref{eq:7}, and \eqref{eq:8} at finite energy to the numerical
calculation from Eq.~\eqref{eq:5}. We see that $B$ is much smaller than
$A$ meaning that $E_0$ is of little significance compared to $\Gamma$.
The variables $\beta$ and $\delta$ are also insignificant, at least a
factor of $10$ smaller than $A$ and $B$. The low-energy dependence of
$\rec$ on energy is thus primarily determined by $K/\Gamma^2$. 
The variations with imaginary strength between $10$
and $120$ amount to only about $10-20$~\%, except for $K$ which decrease
by about a factor of 2. As we will see below, experimental constraints 
limit the imaginary strength variation interval to $\sim 10-25$. 

\begin{figure}[ht!]
  \centering
\begingroup
  \makeatletter
  \providecommand\color[2][]{%
    \GenericError{(gnuplot) \space\space\space\@spaces}{%
      Package color not loaded in conjunction with
      terminal option `colourtext'%
    }{See the gnuplot documentation for explanation.%
    }{Either use 'blacktext' in gnuplot or load the package
      color.sty in LaTeX.}%
    \renewcommand\color[2][]{}%
  }%
  \providecommand\includegraphics[2][]{%
    \GenericError{(gnuplot) \space\space\space\@spaces}{%
      Package graphicx or graphics not loaded%
    }{See the gnuplot documentation for explanation.%
    }{The gnuplot epslatex terminal needs graphicx.sty or graphics.sty.}%
    \renewcommand\includegraphics[2][]{}%
  }%
  \providecommand\rotatebox[2]{#2}%
  \@ifundefined{ifGPcolor}{%
    \newif\ifGPcolor
    \GPcolorfalse
  }{}%
  \@ifundefined{ifGPblacktext}{%
    \newif\ifGPblacktext
    \GPblacktexttrue
  }{}%
  \let\gplgaddtomacro\g@addto@macro
  \gdef\gplbacktext{}%
  \gdef\gplfronttext{}%
  \makeatother
  \ifGPblacktext
    \def\colorrgb#1{}%
    \def\colorgray#1{}%
  \else
    \ifGPcolor
      \def\colorrgb#1{\color[rgb]{#1}}%
      \def\colorgray#1{\color[gray]{#1}}%
      \expandafter\def\csname LTw\endcsname{\color{white}}%
      \expandafter\def\csname LTb\endcsname{\color{black}}%
      \expandafter\def\csname LTa\endcsname{\color{black}}%
      \expandafter\def\csname LT0\endcsname{\color[rgb]{1,0,0}}%
      \expandafter\def\csname LT1\endcsname{\color[rgb]{0,1,0}}%
      \expandafter\def\csname LT2\endcsname{\color[rgb]{0,0,1}}%
      \expandafter\def\csname LT3\endcsname{\color[rgb]{1,0,1}}%
      \expandafter\def\csname LT4\endcsname{\color[rgb]{0,1,1}}%
      \expandafter\def\csname LT5\endcsname{\color[rgb]{1,1,0}}%
      \expandafter\def\csname LT6\endcsname{\color[rgb]{0,0,0}}%
      \expandafter\def\csname LT7\endcsname{\color[rgb]{1,0.3,0}}%
      \expandafter\def\csname LT8\endcsname{\color[rgb]{0.5,0.5,0.5}}%
    \else
      \def\colorrgb#1{\color{black}}%
      \def\colorgray#1{\color[gray]{#1}}%
      \expandafter\def\csname LTw\endcsname{\color{white}}%
      \expandafter\def\csname LTb\endcsname{\color{black}}%
      \expandafter\def\csname LTa\endcsname{\color{black}}%
      \expandafter\def\csname LT0\endcsname{\color{black}}%
      \expandafter\def\csname LT1\endcsname{\color{black}}%
      \expandafter\def\csname LT2\endcsname{\color{black}}%
      \expandafter\def\csname LT3\endcsname{\color{black}}%
      \expandafter\def\csname LT4\endcsname{\color{black}}%
      \expandafter\def\csname LT5\endcsname{\color{black}}%
      \expandafter\def\csname LT6\endcsname{\color{black}}%
      \expandafter\def\csname LT7\endcsname{\color{black}}%
      \expandafter\def\csname LT8\endcsname{\color{black}}%
    \fi
  \fi
  \setlength{\unitlength}{0.0500bp}%
  \begin{picture}(5040.00,3528.00)%
    \gplgaddtomacro\gplbacktext{%
      \csname LTb\endcsname%
      \put(1078,704){\makebox(0,0)[r]{\strut{}$10^{-26}$}}%
      \put(1078,1169){\makebox(0,0)[r]{\strut{}$10^{-24}$}}%
      \put(1078,1635){\makebox(0,0)[r]{\strut{}$10^{-22}$}}%
      \put(1078,2100){\makebox(0,0)[r]{\strut{}$10^{-20}$}}%
      \put(1078,2565){\makebox(0,0)[r]{\strut{}$10^{-18}$}}%
      \put(1078,3030){\makebox(0,0)[r]{\strut{}$10^{-16}$}}%
      \put(4283,484){\makebox(0,0){\strut{}$-10^{2}$}}%
      \put(3379,484){\makebox(0,0){\strut{}$-10^{3}$}}%
      \put(2474,484){\makebox(0,0){\strut{}$-10^{4}$}}%
      \put(1570,484){\makebox(0,0){\strut{}$-10^{5}$}}%
      \put(176,1983){\rotatebox{-270}{\makebox(0,0){\strut{}$\rec$ [cm$^6/$s]}}}%
      \put(2926,154){\makebox(0,0){\strut{}$a [a_0]$}}%
      \put(1930,3141){\makebox(0,0)[r]{\strut{}0 K}}%
      \put(1930,2466){\makebox(0,0)[r]{\strut{}0.5 $\mu$K}}%
      \put(1930,2303){\makebox(0,0)[r]{\strut{}1.0 $\mu$K}}%
      \put(1930,2141){\makebox(0,0)[r]{\strut{}2.0 $\mu$K}}%
      \put(3127,2565){\makebox(0,0){\strut{}$\overbrace{\phantom{sums}}$}}%
      \put(3127,2798){\makebox(0,0){\strut{}$a_C$}}%
    }%
    \gplgaddtomacro\gplfronttext{%
      \csname LTb\endcsname%
      \put(2926,1097){\makebox(0,0)[r]{\strut{}${}^7$Li @ $1.5\mu$K}}%
      \csname LTb\endcsname%
      \put(2926,877){\makebox(0,0)[r]{\strut{}Braaten \& Hammer}}%
    }%
    \gplbacktext
    \put(0,0){\includegraphics{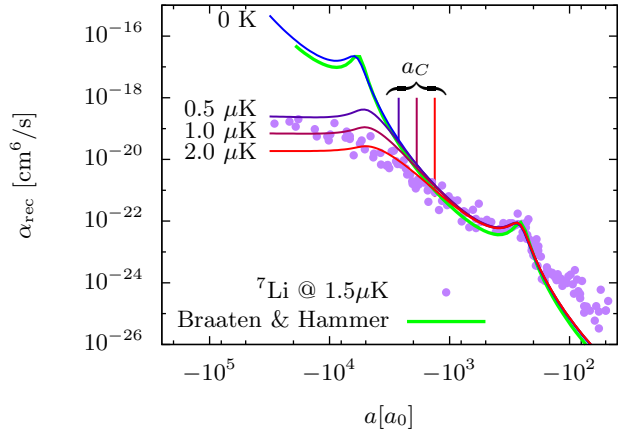}}%
    \gplfronttext
  \end{picture}%
\endgroup
  \caption{Recombination coefficient $\rec$ at zero and finite temperature
    (in $\mu$K) for \Li7 with experimental data at a temperature of
    $1.5\mu$K \cite{hulet2013}. The scattering length, $a$, is in units of
    the Bohr radius, $a_0$. $a_C$ indicates the critical scattering length
    where the height of the barrier equals the mean energy of the atoms.
    At this value the spectrum starts to deviate from the $a^4$ behaviour.}
  \label{figure:4}
\end{figure}

Experiments are performed at finite temperature, as opposed to finite
energy. We therefore average the finite energy calculations using a
normalized Boltzmann distribution for three particles, that is
\begin{equation}
  \langle\rec(a)\rangle_T=\frac{1}{2(k_BT)^3}\int
  E^2e^{-E/k_BT}\rec(a,E)\;dE\;,
  \label{eq:9}
\end{equation}
where $E^2$ arises due to the phase space for three particles. The 
effect of temperature was previously considered in other works such as
Refs.~\cite{dincao2004} and \cite{braaten2008}.
With the
parametrized expression in Eq.~\eqref{eq:6} this folding can readily be
achieved. In the high temperature limit, $T\gg\Gamma$, where only large
energies contribute $\langle\rec(a)\rangle_T\approx\rec(a,E=T\sqrt{2})$.
The opposite limit of very small $T$ naturally leads to
$\langle\rec(a)\rangle_T \approx \rec(a,E \approx 0)$.

\begin{figure}[ht!]
  \centering
\begingroup
  \makeatletter
  \providecommand\color[2][]{%
    \GenericError{(gnuplot) \space\space\space\@spaces}{%
      Package color not loaded in conjunction with
      terminal option `colourtext'%
    }{See the gnuplot documentation for explanation.%
    }{Either use 'blacktext' in gnuplot or load the package
      color.sty in LaTeX.}%
    \renewcommand\color[2][]{}%
  }%
  \providecommand\includegraphics[2][]{%
    \GenericError{(gnuplot) \space\space\space\@spaces}{%
      Package graphicx or graphics not loaded%
    }{See the gnuplot documentation for explanation.%
    }{The gnuplot epslatex terminal needs graphicx.sty or graphics.sty.}%
    \renewcommand\includegraphics[2][]{}%
  }%
  \providecommand\rotatebox[2]{#2}%
  \@ifundefined{ifGPcolor}{%
    \newif\ifGPcolor
    \GPcolorfalse
  }{}%
  \@ifundefined{ifGPblacktext}{%
    \newif\ifGPblacktext
    \GPblacktexttrue
  }{}%
  \let\gplgaddtomacro\g@addto@macro
  \gdef\gplbacktext{}%
  \gdef\gplfronttext{}%
  \makeatother
  \ifGPblacktext
    \def\colorrgb#1{}%
    \def\colorgray#1{}%
  \else
    \ifGPcolor
      \def\colorrgb#1{\color[rgb]{#1}}%
      \def\colorgray#1{\color[gray]{#1}}%
      \expandafter\def\csname LTw\endcsname{\color{white}}%
      \expandafter\def\csname LTb\endcsname{\color{black}}%
      \expandafter\def\csname LTa\endcsname{\color{black}}%
      \expandafter\def\csname LT0\endcsname{\color[rgb]{1,0,0}}%
      \expandafter\def\csname LT1\endcsname{\color[rgb]{0,1,0}}%
      \expandafter\def\csname LT2\endcsname{\color[rgb]{0,0,1}}%
      \expandafter\def\csname LT3\endcsname{\color[rgb]{1,0,1}}%
      \expandafter\def\csname LT4\endcsname{\color[rgb]{0,1,1}}%
      \expandafter\def\csname LT5\endcsname{\color[rgb]{1,1,0}}%
      \expandafter\def\csname LT6\endcsname{\color[rgb]{0,0,0}}%
      \expandafter\def\csname LT7\endcsname{\color[rgb]{1,0.3,0}}%
      \expandafter\def\csname LT8\endcsname{\color[rgb]{0.5,0.5,0.5}}%
    \else
      \def\colorrgb#1{\color{black}}%
      \def\colorgray#1{\color[gray]{#1}}%
      \expandafter\def\csname LTw\endcsname{\color{white}}%
      \expandafter\def\csname LTb\endcsname{\color{black}}%
      \expandafter\def\csname LTa\endcsname{\color{black}}%
      \expandafter\def\csname LT0\endcsname{\color{black}}%
      \expandafter\def\csname LT1\endcsname{\color{black}}%
      \expandafter\def\csname LT2\endcsname{\color{black}}%
      \expandafter\def\csname LT3\endcsname{\color{black}}%
      \expandafter\def\csname LT4\endcsname{\color{black}}%
      \expandafter\def\csname LT5\endcsname{\color{black}}%
      \expandafter\def\csname LT6\endcsname{\color{black}}%
      \expandafter\def\csname LT7\endcsname{\color{black}}%
      \expandafter\def\csname LT8\endcsname{\color{black}}%
    \fi
  \fi
  \setlength{\unitlength}{0.0500bp}%
  \begin{picture}(5040.00,3528.00)%
    \gplgaddtomacro\gplbacktext{%
      \csname LTb\endcsname%
      \put(1078,968){\makebox(0,0)[r]{\strut{}$10^{-24}$}}%
      \put(1078,1496){\makebox(0,0)[r]{\strut{}$10^{-22}$}}%
      \put(1078,2023){\makebox(0,0)[r]{\strut{}$10^{-20}$}}%
      \put(1078,2551){\makebox(0,0)[r]{\strut{}$10^{-18}$}}%
      \put(1078,3079){\makebox(0,0)[r]{\strut{}$10^{-16}$}}%
      \put(4643,484){\makebox(0,0){\strut{}$-10^{2}$}}%
      \put(3715,484){\makebox(0,0){\strut{}$-10^{3}$}}%
      \put(2787,484){\makebox(0,0){\strut{}$-10^{4}$}}%
      \put(1859,484){\makebox(0,0){\strut{}$-10^{5}$}}%
      \put(176,1983){\rotatebox{-270}{\makebox(0,0){\strut{}$\rec$ [cm$^6/$s]}}}%
      \put(2926,154){\makebox(0,0){\strut{}$a [a_0]$}}%
      \put(1915,3079){\makebox(0,0)[r]{\strut{}0 K}}%
      \put(1915,2735){\makebox(0,0)[r]{\strut{}1 nK}}%
      \put(1915,2471){\makebox(0,0)[r]{\strut{}4 nK}}%
      \put(1915,2160){\makebox(0,0)[r]{\strut{}15 nK}}%
      \put(1915,1833){\makebox(0,0)[r]{\strut{}60 nK}}%
      \put(1915,1506){\makebox(0,0)[r]{\strut{}240 nK}}%
    }%
    \gplgaddtomacro\gplfronttext{%
      \csname LTb\endcsname%
      \put(3656,3090){\makebox(0,0)[r]{\strut{}853.07 G}}%
      \csname LTb\endcsname%
      \put(3656,2870){\makebox(0,0)[r]{\strut{}554.71 G}}%
      \csname LTb\endcsname%
      \put(3656,2650){\makebox(0,0)[r]{\strut{}553.3 G}}%
    }%
    \gplbacktext
    \put(0,0){\includegraphics{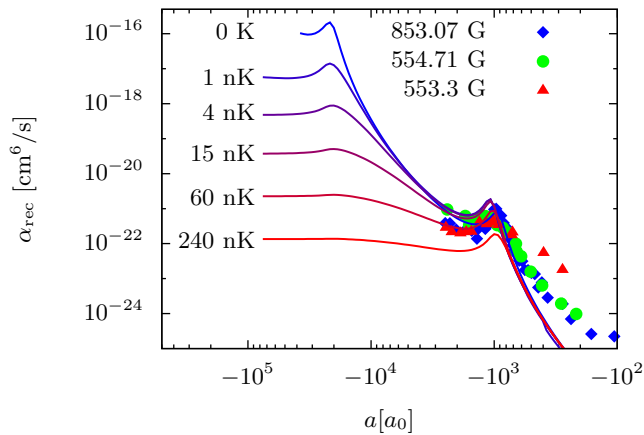}}%
    \gplfronttext
  \end{picture}%
\endgroup
  \caption{Recombination coefficient $\rec$ at zero and finite temperature
    for \Cs~with experimental data taken at a temperature of about
    $15$~nK \cite{berninger2011}.}
  \label{figure:5}
\end{figure}

We now compare experimental data and numerical calculations with
our imaginary potential. Full numerical results and the parametrization 
are virtually indistinguishable. The experimental recombination 
data for \Li7 \cite{hulet2013} along with calculations from our
model at zero and finite temperature is shown in \fig{figure:4}. 
The only pronounced measured peak at
$a \approx -280 a_0$ (where $a_0$ is the Bohr radius) 
is well described by our model. The peak
position is fitted with $\rim=0.41 a_0$ and the overall shape of the peak is
fitted with $\Vim=-68\hbar^2/ma_{0}^{2}$. For zero temperature we find for all $a$ almost
precisely the same as the zero-energy formula of Ref.~\cite{bra2006} where $\eta^-=0.12$
and $a^{(-)}=-241a_0$ \cite{hulet2013}. At finite temperature, we find the observed lowering of
recombination rates for large $a$. This flattening of $\rec$ appears for
temperatures exceeding the barrier height, in other words for
$a^2>a_C^2\equiv0.143\hbar^2a_0^2/(mTk_B)$ as is shown in \fig{figure:4}
for the temperatures indicated.
Recombination rates are also measured for \Cs~at $T\sim 15$~nK for three
different Feshbach resonances \cite{berninger2011} which show very similar
behavior. These are shown in \fig{figure:5} along with our calculations
for different temperatures using $\rim=1.58 a_0$ and $\Vim=-10\hbar^2/ma_{0}^{2}$.
Our model reproduces the data for all three resonances with the same model parameters.
No data exists at $-2\times10^4a_0$ where we predict another peak. 
From \fig{figure:5} we conclude that a
temperature below $\sim2$~nK seems to be required to observe this peak.

\begin{figure}[ht!]
  \centering
\begingroup
  \makeatletter
  \providecommand\color[2][]{%
    \GenericError{(gnuplot) \space\space\space\@spaces}{%
      Package color not loaded in conjunction with
      terminal option `colourtext'%
    }{See the gnuplot documentation for explanation.%
    }{Either use 'blacktext' in gnuplot or load the package
      color.sty in LaTeX.}%
    \renewcommand\color[2][]{}%
  }%
  \providecommand\includegraphics[2][]{%
    \GenericError{(gnuplot) \space\space\space\@spaces}{%
      Package graphicx or graphics not loaded%
    }{See the gnuplot documentation for explanation.%
    }{The gnuplot epslatex terminal needs graphicx.sty or graphics.sty.}%
    \renewcommand\includegraphics[2][]{}%
  }%
  \providecommand\rotatebox[2]{#2}%
  \@ifundefined{ifGPcolor}{%
    \newif\ifGPcolor
    \GPcolorfalse
  }{}%
  \@ifundefined{ifGPblacktext}{%
    \newif\ifGPblacktext
    \GPblacktexttrue
  }{}%
  \let\gplgaddtomacro\g@addto@macro
  \gdef\gplbacktext{}%
  \gdef\gplfronttext{}%
  \makeatother
  \ifGPblacktext
    \def\colorrgb#1{}%
    \def\colorgray#1{}%
  \else
    \ifGPcolor
      \def\colorrgb#1{\color[rgb]{#1}}%
      \def\colorgray#1{\color[gray]{#1}}%
      \expandafter\def\csname LTw\endcsname{\color{white}}%
      \expandafter\def\csname LTb\endcsname{\color{black}}%
      \expandafter\def\csname LTa\endcsname{\color{black}}%
      \expandafter\def\csname LT0\endcsname{\color[rgb]{1,0,0}}%
      \expandafter\def\csname LT1\endcsname{\color[rgb]{0,1,0}}%
      \expandafter\def\csname LT2\endcsname{\color[rgb]{0,0,1}}%
      \expandafter\def\csname LT3\endcsname{\color[rgb]{1,0,1}}%
      \expandafter\def\csname LT4\endcsname{\color[rgb]{0,1,1}}%
      \expandafter\def\csname LT5\endcsname{\color[rgb]{1,1,0}}%
      \expandafter\def\csname LT6\endcsname{\color[rgb]{0,0,0}}%
      \expandafter\def\csname LT7\endcsname{\color[rgb]{1,0.3,0}}%
      \expandafter\def\csname LT8\endcsname{\color[rgb]{0.5,0.5,0.5}}%
    \else
      \def\colorrgb#1{\color{black}}%
      \def\colorgray#1{\color[gray]{#1}}%
      \expandafter\def\csname LTw\endcsname{\color{white}}%
      \expandafter\def\csname LTb\endcsname{\color{black}}%
      \expandafter\def\csname LTa\endcsname{\color{black}}%
      \expandafter\def\csname LT0\endcsname{\color{black}}%
      \expandafter\def\csname LT1\endcsname{\color{black}}%
      \expandafter\def\csname LT2\endcsname{\color{black}}%
      \expandafter\def\csname LT3\endcsname{\color{black}}%
      \expandafter\def\csname LT4\endcsname{\color{black}}%
      \expandafter\def\csname LT5\endcsname{\color{black}}%
      \expandafter\def\csname LT6\endcsname{\color{black}}%
      \expandafter\def\csname LT7\endcsname{\color{black}}%
      \expandafter\def\csname LT8\endcsname{\color{black}}%
    \fi
  \fi
  \setlength{\unitlength}{0.0500bp}%
  \begin{picture}(5040.00,3528.00)%
    \gplgaddtomacro\gplbacktext{%
      \csname LTb\endcsname%
      \put(946,704){\makebox(0,0)[r]{\strut{}$10^{-9}$}}%
      \put(946,988){\makebox(0,0)[r]{\strut{}$10^{-8}$}}%
      \put(946,1273){\makebox(0,0)[r]{\strut{}$10^{-7}$}}%
      \put(946,1557){\makebox(0,0)[r]{\strut{}$10^{-6}$}}%
      \put(946,1842){\makebox(0,0)[r]{\strut{}$10^{-5}$}}%
      \put(946,2126){\makebox(0,0)[r]{\strut{}$10^{-4}$}}%
      \put(946,2411){\makebox(0,0)[r]{\strut{}$10^{-3}$}}%
      \put(946,2695){\makebox(0,0)[r]{\strut{}$10^{-2}$}}%
      \put(946,2980){\makebox(0,0)[r]{\strut{}$10^{-1}$}}%
      \put(946,3264){\makebox(0,0)[r]{\strut{}$10^{0}$}}%
      \put(4257,484){\makebox(0,0){\strut{}$-10^{2}$}}%
      \put(2973,484){\makebox(0,0){\strut{}$-10^{3}$}}%
      \put(1690,484){\makebox(0,0){\strut{}$-10^{4}$}}%
      \put(176,1983){\rotatebox{-270}{\makebox(0,0){\strut{}$\gamma$}}}%
      \put(2860,154){\makebox(0,0){\strut{}$a [a_0]$}}%
      \put(1889,2044){\makebox(0,0)[r]{\strut{}76.8 $\mu$K}}%
      \put(1889,1873){\makebox(0,0)[r]{\strut{}38.4 $\mu$K}}%
      \put(1889,1702){\makebox(0,0)[r]{\strut{}19.2 $\mu$K}}%
      \put(1889,1530){\makebox(0,0)[r]{\strut{}9.6 $\mu$K}}%
      \put(1889,1358){\makebox(0,0)[r]{\strut{}4.8 $\mu$K}}%
      \put(1889,1188){\makebox(0,0)[r]{\strut{}2.4 $\mu$K}}%
      \put(1889,1014){\makebox(0,0)[r]{\strut{}1.2 $\mu$K}}%
      \put(1889,864){\makebox(0,0)[r]{\strut{}0.6 $\mu$K}}%
    }%
    \gplgaddtomacro\gplfronttext{%
    }%
    \gplbacktext
    \put(0,0){\includegraphics{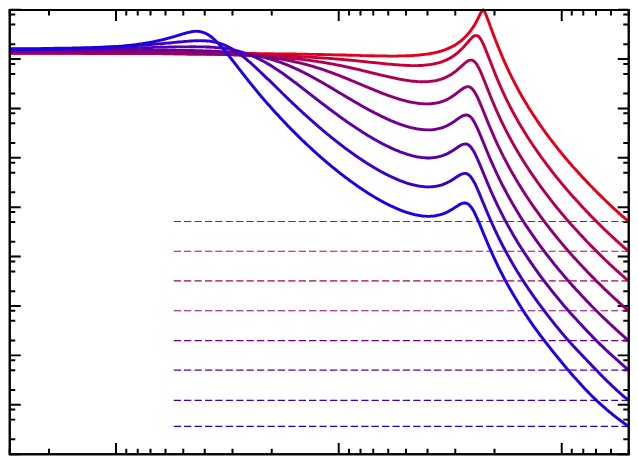}}%
    \gplfronttext
  \end{picture}%
\endgroup
  \caption{The decay parameter $\gamma$ for the \Li7 system as a function of scattering length
  for finite energies in temperature units. At high $|a|$ all curves have
  about the same value $\sim0.14$, independent of energy.}
  \label{figure:6}
\end{figure}

Increasing $|a|$ moves the barrier towards infinity and reduces $E_B(a)$,
which is thus exceeded by typical laboratory temperatures. This means
that the high-energy limit is approached and an $a$-independent
recombination rate is obtained. The energy dependence in this limit is
$1/E^2$ and the value of $K$ determines the limiting
value of $\rec$. We show in \fig{figure:6} the calculated values of
$\gamma$ as a function of $a$ for several finite energies. All
$\gamma$-values are lowered but for large $|a|$ an energy and
$a$~independent constant of~$\approx0.14$ is reached. This value
depends on the strength of the imaginary potential $m\Vim\rim^2/\hbar^2$, which
controls the height and shape of the absorption as function of both $E$ and $a$.
This value is numerically deceivingly similar to the $\eta^-$ of
\cite{hulet2013} used to fit the peak in \fig{figure:4}. Formally there is
also a connection although $\eta^-$ is more complicated
and derived through multiple scattering theory for zero energy \cite{bra2006}. The
physical meaning is different from our $\gamma$ and the expressions are
not one-to-one related.

In conclusion, we present a simple and physically transparent model of three-body
recombination for negative scattering lengths that does not require 
a short-range three-body cut-off. Instead it includes an imaginary potential
at short distance that takes decay into deeply bound dimers into account.
Full numerical solution of the three-body equations were used to obtain
the recombination rate and subsequently a parametrization in terms of 
the Breit-Wigner resonance formula was presented and shown to display 
the expected scaling behavior. Finally we showed how this new model reproduces
experimental data on \Li7 and \Cs. If we express the radius of the imaginary
potential in units of the van der Waals length we find $\rim/r_\textrm{vdW}=0.0063$
and $\rim/r_\textrm{vdW}=0.0078$ respectively, while the strength
is $|\Vim|/V_\textrm{vdW}=2.87\cdot 10^5$ and $|\Vim|/V_\textrm{vdW}=4.08\cdot 10^5$ where $V_\textrm{vdW}=\hbar^2/mr_\textrm{vdW}^2$. 
The similarity of $\rim$ and $\Vim$ in van der Waals units
indicates that there could be universality hidden in this
parameter. The differences that we find is most likely related to
the difference in deeply bound state in the two-body potentials of 
\Li7 and \Cs.
Further studies using models that include realistic potentials
for the deep dimers are needed to fix the parameters of our model to
state-of-the-art short-range calculations and data.

\end{document}